\documentclass[journal]{IEEEtran}
\ifCLASSINFOpdf
  \usepackage[pdftex]{graphicx}
  \graphicspath{{.}{../pdf/}{../jpeg/}}
  \DeclareGraphicsExtensions{.pdf,.jpeg,.jpg,.png,.gif}
\else
\fi
\begin{document}
%
\title{A Novel RTL ATPG Model Based on\\Gate Inherent Faults (GIF-PO) of Complex Gates}
%
%
%

\author{Tobias~Strauch
\thanks{Manuscript published December 15th, 2016.}}

%
%

\markboth{T. Strauch: A Novel RTL ATPG Model Based on Gate Inherent Faults (GIF-PO) of Complex Gates}%
{Shell \MakeLowercase{\textit{et al.}}: Bare Demo of IEEEtran.cls for IEEE Journals}
%



\maketitle

\begin{abstract}
This paper starts with a comprehensive survey on RTL ATPG. It then proposes a novel RTL ATPG model based on ''Gate Inherent Faults'' (GIF). These GIF are extracted from each complex gate (adder, case-statement, etc.) of the RTL source code individually. They are related to the internal logic paths of a complex gate. They are not related to any net/signal in the RTL design. It is observed, that when all GIF on RTL are covered (100\%) and the same stimulus is applied, then all gate level stuck-at faults of the netlist are covered (100\%) as well. The proposed RTL ATPG model is therefore synthesis independent. This is shown on ITC'’99 testcases. The applied semi-automatic test pattern generation process is based on functional simulation.
\end{abstract}

\begin{IEEEkeywords}
Survey on RTL ATPG, high-level test generation, register transfer level (RTL) test generation, gate inherent faults
\end{IEEEkeywords}

%
\IEEEpeerreviewmaketitle

\section{Introduction}
Automatic test pattern generation (ATPG) on gate level for production tests (stuck-at) is well accepted and understood. Using register transfer level (RTL) descriptions of the design to generate or support the generation of gate level test sets has been discussed in various publications over several the last decades. This paper begins with a comprehensive survey of publications on the subject of RTL ATPG.\\
\indent The goal is to reach the highest possible gate level coverage, preferable 100\% stuck-at coverage. One desirable scenario is to generate a test set, which is generated based on the RTL source code, and which achieves 100\% gate level coverage on any correlating netlist, independently of the synthesis result. A post processing optimization of the test set based on a concrete implementation is acceptable. The author is not aware of such a published solution.\\
\indent A methodology is proposed, which is based on a ''Gate Inherent Faults to Primary Outputs'' (GIF-PO) model. RTL code can be seen as a high level netlist based on complex gates like adders, case-statements etc. The individual gate inherent faults for each one of these complex gates can be automatically extracted. The RTL code achieves 100\% RTL coverage, if (and only if) a test set propagates all GIF to all relevant primary outputs (registers, design outputs) in order to be captured. It is observed, that in this case, the resulting netlist achieves 100\% gate level stuck-at coverage, independently of the individual synthesis result. The test set can then be further reduced based on a given gate level netlist.\\
\indent RTL ATPG models and methodologies can be used for standard, partial and non-scan designs, etc.. They can support the generation of single capture cycles as well as multiple capture cycle based test pattern (''sequential ATPG''). The pattern generation process can be done solely based on the RTL design or with the help of functional tests. Test sets can be generated fully automatically or with user interaction (semi-automatic, ''semi-ATPG'').\\
\indent It is outside the scope of this paper to apply the proposed GIF-PO coverage model to all these different kinds of solutions. This paper concentrates on using the GIF-PO coverage model on semi-automatic (user interactive) ATPG based on functional tests. This is done to outline the concept, how the numbers reported in the result section are generated. Subsequent work will be required to combine the GIF-PO coverage model with more sophisticated functional pattern generation techniques.\\ 
\indent This paper is organized as follows. In Section 2 a comprehensive survey on RTL ATPG is given. Section 3 introduces the GIF-PO coverage model. A semi-ATPG framework is shown in Section 4 that also outlines, how the numbers in the final result Section are generated.

\section{Survey on RTL ATPG}
\indent Harris gives a comprehensive overview on RTL ATPG in 2001 [1], in particular on the usage of control/datapath flow graphs (CDFG). This section gives an overview of the different RTL fault models for testability analysis on RTL and/or test pattern generation not mentioned in [1]. Solutions which propose additional test insertion logic are not considered. The individual publications are only mentioned once, although many of them fall into multiple categories. The survey starts with work that utilizes functional RTL tests in one way or the other to generate gate level stuck-at tests, because the semi-ATPG flow used in this paper falls into this category.

\subsection{Related work on functional simulation based RTL ATPG}
\indent The following related work can be found regarding the combination of functional RTL test simulation and gate level structural coverage. A functional coverage metric to estimate the gate level fault coverage of functional tests has been outlined by Patil et al. in [2]. They borrow concepts from simulation-based design verification by defining fault detection conditions as coverage objects and monitoring their occurrence, also called their hit counts, during RTL simulation. Fang et al. propose in [3, 4] output deviations as a metric to grade functional test sequences at the RTL without explicit fault simulation.\\
\indent Hobeika et al. establish in [5] the relationship between existing dark corners in functional verification and hard to detect stuck-at faults. Based on this relation, they explore the use of structural test patterns in the verification process and compare the results to well-known verification techniques.\\
\indent Sanyal et al. show in [6], that their proposed method significantly increases the gate level defect coverage of existing functional test sequences (on RTL) by selecting a small set of control and observation points.\\
\indent Mao et al. outline in [7] an RTL fault grading mechanism, which takes Verilog RTL models and functional verification patterns as the inputs, and generates a quantitative RTL fault coverage. It is also able to provide information for test pattern improvement to increase fault coverage and design modifications to improve testability at RTL.\\
\indent Santos et al. [8] propose an RTL based defect oriented (DO) test generation methodology, for which a high defects coverage can be derived. It uses RTL fault simulation to identify dark corners and to guide the identification of partially constrained test vectors, which may significantly increase the single or multiple detection of RTL faults in these dark corners. Multiple RTL fault detection will increase the probability of detection of defects on the synthesized structure.\\
\indent Zhang et al. use automatic test instruction generation (ATIG) based on processor self-testing algorithms and expanded instructions to generate sequential test sets for gate level faults in [9].\\
\indent An automatic functional test pattern generator on RTL is used by Ferrandi et al. in [10]. The adopted functional test pattern generator is based on implicit techniques (e.g. based on Binary Decision Diagrams) that compare the erroneous and error-free VHDL descriptions to identify (if possible) a functional test pattern. Such a test set is then tailored to the particular gate-level implementation by transforming it into a specific test sequence, based on the scheduling adopted by the high-level synthesis.
\subsection{Fault models based on 9-Value algebra}
\indent Chaiyakul et al. propose assignment decision diagrams (ADD) for logic synthesis in [11]. Ghosh et al. enhanced this modeling technique in [12, 13, 14, 15] by a nine value algebra for RTL ATPG. Ghosh et al. then improve this method by using observability-enhanced tag coverage [16]. Lingappan et al. enhance the ADD based test generation technique by a satisfiability based approach in [17, 18] and optimize it for microprocessors in [19].\\
\indent Obien et al. [20] use ADD based nine value algebra  and combine it with pseudo primary-inputs and pseudo primary-outputs to generate constrained ATPG on RTL.
\subsection{Control and/or datapath oriented RTL fault models}
\indent Lingappan et al. also enhance the ADD based test generation technique by control-data flow graphs (CDFG) and state transition sequence analysis in [21, 22]. Further on, the technique is enhanced by Boolean implications (also known as as the unsatisfiable segment) in [23, 24].\\
\indent Chen et al. [25] convert the behavioral description given in the form of a Control Flow Graph (CFG). During the path analysis, symbolic implication is used to reduce the number of temporary variables. After the path analysis, a variable classification is used to explore the intrinsic controllability and observability of the circuit.\\
\indent Gosh et al. propose a technique in [26, 27], that analyzes the data path and controller of an RTL circuit and extracts a test control-data flow (TCDF) graph. This graph is then used to test the circuit hierarchically by symbolically justifying and propagating precomputed test sets of modules, registers and multiplexers in the circuit from system inputs and propagating the output responses to system outputs.\\
\indent Lee et al. outline in [28] ATPG for microprocessors based on their behavioral knowledge, which is modeled by structural data-flow graphs (SDG).\\
\indent Raik and Ubar propose in [29, 30] that an RTL model of the control and datapath parts is represented by high-level (word-level) decision diagrams (DD), whereas the gate-level descriptions of RTL blocks are given by structurally synthesized binary decision diagrams. The method proposed in the paper combines deterministic and simulation-based techniques for test pattern generation.\\
\indent Corno et al. propose in [31] a technique, where the RTL circuit description is automatically analyzed to extract static structural information, control and data dependencies, and to group statements in basic-blocks. Then a code coverage approach is exploited to excite the RT-level assignment single-bit faults. After excitation, fault effect propagation and observation are tackled utilizing simulation scripts. In conclusion, a fault dropping phase is run to optimize the process.\\
\indent Wada et al. propose in [32] a DFT method to modify RTL data paths so that a test plan for every module can be generated with lower complexity. It is based on a multiplexed data path architecture.\\
\indent Fallah et al. present in [33] a new hardware description language (HDL) SAT checking algorithm that works directly on the HDL model. The primary feature of the algorithm is a seamless integration of linear-programming techniques for feasibility checking of arithmetic equations, that govern the behavior of data-path modules, and SAT checking for logic equations, that govern the behavior of control modules.\\
\indent Yin et al. show in [34] how process controlling trees (PCT) and data dependence graphs (DDG) are automatically extracted from RTL description to guide a simulation based ATPG method.\\
\indent Yadavalli et al. discuss in [35] an automatic test scheduling system for architectures that use separate control- and datapaths. It provides a system with eleven signal types to perform test scheduling at the RTL which allows module level precomputed test sets to be directly used for gate level testing.
\subsection{Fault models based on low level transformation}
\indent Dave et. al. propose in [36] to represent a circuit by an equivalent two-level AND-OR circuit. Their work suggests the use of stuck type faults in the two-level AND-OR equivalent of a combinational function as an adequate set of faults for test generation. Jacob et al. optimize this approach in [37] for finite state machines (FSM) and sequential tests.\\
\indent Corno et al adopt in [38] a particular instantiation of the observability enhanced statement coverage metric. In particular, single stuck-at bit faults are modeled on all assignment targets of the executed statements that respect a defined set of rules. 
\subsection{Fault models based on input/ouput relations}
\indent Pomeranz et al. define in [39] a fault coverage metric, called the PI-PO fault coverage, based on stuck-at faults on primary inputs of an RTL logic block. It is used to estimate the gate level stuck-at fault coverage of a test set for this block.\\
\indent Kang et al. introduce in [40] an input/output transition fault-coverage metric at the register-transfer. This coverage metric can be used in evaluating functional tests for high volume manufacturing as well as in early testability analysis.
\subsection{Fault models based on hierarchical structure}
\indent Jervan et al. assess the effectiveness of high-level test generation with a simple ATPG algorithm in [41], and present a novel high-level hierarchical test generation (HTG) approach to improve the results obtained by a pure high-level test generator.\\
\indent Safert et al. show in [42, 43] the advantages of exploiting both the hierarchy in circuit description and the knowledge concerning the function of so-called highlevel primitives (HLP’s) during ATG. It leads to significant improvements in implication, unique sensitization, and multiple backtrace, all of which play a key role in the efficiency of any ATPG system.\\
\indent Vedula et al. propose in [44] a hierarchical test generation method by using functional constraint extraction/composition based on VHDL RTL.\\
\indent Lee et al. address in [45] both top-down and bottom-up approaches in hierarchical test generation. It is shown that the valid control code abstraction and test cube justification techniques are very effective to overcome the architectural level functional constraint problem and to achieve high efficiency in test computation.\\
\indent Makris et al. introduce in [46, 47] a formal mechanism for capturing test justification and propagation related behavior of blocks. Based on the identified test translation behavior, an RTL testability analysis methodology for hierarchical designs is derived. An algorithm for pinpointing the local-to-global test translation controllability and observability bottlenecks is presented. Makris et al. refine the concept of modular transparency by emphasizing on the channel notion in [48, 49], which is a powerful mechanism that captures modular transparency in terms of bijection functions defined on variable bitwidth signal entities.\\
\indent Kunda et al. present a technique in [50] for test vector generation based on high-level primitives and a general fault model. The technique is able to handle circuits in a hierarchical fashion, and treats the signals at a bit-vector level rather than at the bit level.\\
\indent Magdolen et al. show in [51] a TPG system (REGGEN) that can generate test set for RTL models. The basic TPG technique on RT-level was extended by fault simulation process and new proposed simplifying rules for individual operators were implemented.
\subsection{State machine fault models}
\indent Li et al. show in [52] that states of a FSM, or clusters of states, which are termed behavioral phases in high-level descriptions, can be obtained through the analysis of behavioral descriptions as well as functional descriptions. Clustering of circuit states or behavioral phases reduces the complexity of analysis on state space, which can represent the function of a circuit more explicitly and refinedly. A refined representation is used in the test generation algorithm to simplify and speed up search process of test sequences.\\
\indent Hosokawa et al. propose in [53, 54] both a fault-independent test generation method and a fault-dependent test generation method for state observable FSMs.\\
\indent Guglielmo et al. [55] propose RTL ATPG that explores the design under test (DUT) state space by exploiting an easy-to-traverse extended finite state machine (EFSM). The ATPG engine relies on learning, back-jumping and constraint logic programming to deterministically generate test vectors for traversing all transitions of the extended FSM.\\
\indent Mirzaei et al. use a hybrid canonical data structure based on a Horner expansion diagram (HED) in [56] to generate test patterns from RTL.\\
\indent Pomeranz et al. propose in [57] a functional fault model to generate sequential tests with very high gate-level stuck-at fault coverage. Their goal is to show that a functional fault model does not have to include fault effect propagation or observability requirements. Instead, it is possible to define conditions to be satisfied by a test sequence based only on the fault free circuit.\\
\indent Lajolo et al. address the POLIS co-design environment [58], and developed a fault model that mimics the permanent single stuck-at fault model while at the behavioral level. In POLIS the system is represented as a network of interacting Codesign Finite State Machines (CFSMs). CFSMs extend Finite State Machines with arithmetic computations without side effects on transition edges.
\subsection{Considering gate level fault models}
\indent Ravi et al. show an RTL to gate level correspondence in [59] and RTL versus gate level coverage correlation is elaborated on by Rumplik et al. in [60].\\
\indent Al-Yamani et al. compare in [61] the test-quality based on test sets obtained by complex (high-level) gates and test sets obtained by elementary gates. The paper claims that there is a significant penalty in test quality for using complex gates (structural RTL) as fault sites with the single-stuck model.\\
\indent Rudnick et al. propose in [62] software testing based techniques at the high level combined with test enhancement techniques at the gate level by identifying the datapath and control portions of the description and by using state transition graphs (STG) for the control machine.\\
\indent Vinnakota et al. show in [63] how gate-level faults are translated to functional faults and test generation is performed at the functional level. The techniques were incorporated into a multilevel test pattern generation algorithm for sequential logic circuits.\\
\indent Pradhu et al. [64] map the gate level stuck-at fault to RTL and build an equivalent faulty RTL model. The fault activation and propagation constraints are captured using control and data flow graph of RTL as an liner temporal logic (LTL) property. This LTL property is then negated and given to a bounded model checker based on a bit-vector satisfiability module theories (SMT) solver.\\
\indent Ravi et al. [65] propose a simple two-pass strategy that couples register-transfer level (RTL) test generation with gate-level sequential test generation through fault lists. They motivate this approach by showing that faults found hard-to-test by gate-level sequential test generators are often easily testable at the RTL. Likewise, modules found symbolically untestable at the RTL have many of their faults testable at the gate level. Therefore, a two-pass strategy, which runs a fast RTL test generator followed by a gate-level sequential test generator on the remaining untested faults, can leverage off the strengths of each test generator.\\
\indent Chen et al. [66] demonstrate a methodology which uses gate-level and architecture information to improve coverage for structural faults. This software based self-testing methodology uses an automatic test pattern generation tool to generate the constrained test patterns to effectively test the combinational fundamental intellectual properties used in the processor. The approach refers to the RTL code and processor architecture for the rest of the control and steering logic for test routine development.
\subsection{RTL ATPG using fault/error injection}
\indent Thaker et al. use in [67] stratified fault sampling  in RTL fault simulation to estimate the gate-level fault coverage of given test patterns. RTL fault modeling and fault-injection algorithms are developed such that the RTL fault list of a module can be treated as a representative fault sample of the collapsed gate-level stuck-at fault set of the module.\\
\indent Maniatakos et al. analyze in [68] the impact of RTL and gate level faults on the instruction execution flow of a microprocessor, for which fault-simulating the entire gate level model would probably be impractical. Fault simulation is done using error injection on RTL and gate level.\\
\indent Camos et al. [69] introduce a general description of abstract mutation based design error models that can be tailored to span any coverage measure for microprocessor validation. They then present a method of effectively using this statistical information to guide the ATPG efforts.
\subsection{Textual fault models}
\indent Hobeika et al. propose in [70, 71] an RTL illegal state extraction methodology based on code parsing and parsing expression grammar (PEG). The extracted values help building functional constraints that are forced during ATPG process in order to generate pseudo-functional test patterns.\\
\indent Chen et al. present in [72] a complete behavioral fault simulation and ATPG system for circuits modeled in VHDL. Ten different HDL behavioral fault models are selected and used to generate test patterns through fault simulation.
\subsection{Knowledge based RTL ATPG}
\indent Pecenka et al. have developed a method [73] which utilizes controllability and observability parameters to estimate the overall circuit testability on RTL. A simple evolutionary algorithm is used for developing a circuit with required properties. The initial population consisting of P individuals is generated randomly. New populations are formed using tournament selections and mutation operators.\\
\indent Lynch et al. present in [74] an approach for using the RTL description of a circuit together with artificial intelligence (AI) concepts in a diagnostic ATPG system for gate level stuck-at faults. The AI based concepts are implemented using genetic algorithms and adaptive resonance theory neural network.\\
\indent Various testability metrics are considered by Corno et al. in [75, 76], whereas the test pattern generation phase is combined with genetic algorithms. Their proposed metrics are borrowed from software testing and are for instance branch coverage, different values, observation tree and observation token based metrics.\\
\indent Vishakantaiah et al. introduce in [77] techniques to automatically generate test knowledge from structural and behavioral information in the VHDL description of a design by using a module operation tree.\\
\indent Pitchumani et al. show in [78] how they use functional models of VHDL constructs and its correspondence between a faulty VHDL construct and an equivalent field replaceable unit. This is used for fault diagnosis by constraint suspension. Constraint suspension is a way of diagnosing faults by reasoning from first principles using knowledge of the structure and behavior of the circuit.\\
\indent Santos et al. introduce in [79] an RTL testability metric called implicit functionality multiple branch (IFMB) metric that evaluates the exercise of implicit functionality and multiple branch coverage of conditional constructs. It is claimed that it leads to higher correlation values between the RTL faults and the realistic faults detection.\\
\indent Ohtake et al. [80] define novel test knowledge extracted method from RTL description and propose a method for test generation of weakly testable data paths using the test knowledge. The test knowledge estimated by the proposed heuristic measures can reduce the search space of sequential ATPG tools effectively.
\subsection{RTL ATPG and TLM}
\indent Javaheri et al. [81] use transaction level modeling (TLM) and communication hardware. They propose a set of high-level fault models that include faults for data and control parts of a communication link. They show how the proposed high-level fault models map into faults at the gate level in communication hardware.
\subsection{BIST RTL ATPG}
Lin et al. [82] show a pseudofunctional-test methodology that attempts to minimize the overtesting problem of the scan-based circuits in automatic test pattern generation (ATPG) and built-in self-test (BIST) test generation approaches. The first pattern of a two-pattern test is still delivered by scan in the test mode but the pattern is generated in such a way that it does not violate the functional constraints extracted from the functional logic. The second pattern is then generated in a functional mode using the functional justification (also called broadside) test application scheme.\\
\indent Tsai et al. [83, 84] present two techniques for improving the test quality of existing scan-based BIST architectures. They first propose an almost-full-scan BIST solution in which they identify and remove a small number of scan flip-flops from the scan chains to achieve a higher FC with shorter test application time comparing to the full-scan circuit. They then present a general test application scheme for scan-based BIST which employs multiple test sessions with a unique number of capture cycles in each test session.\\
\indent Santos et al. [85] propose mask-based BIST TPG improvements, namely in two areas: RTL estimation of the test length to be used for each mask, in order to reach high defects coverage, and the identification of an optimum mask for each set of nested RTL conditions.\\
\indent Berthelot et al describe in [86] a new method for implementing a test-per-clock BIST scheme for operators in datapaths. The method targets minimal area overhead. It aims at BISTing datapaths as early as possible in the design flow. The FC to achieve on datapath operators is a user-given parameter; it is used to guide the choice of a TPG among all possibilities. Finally, several types of TPGs/CMPs can be inserted within the same design for better optimization.\\
\indent Masuzawa et al. [87] present a new BIST method for RTL data paths based on single-control testability, a new concept of testability. The BIST method adopts hierarchical test. Test pattern generators are placed only on primary inputs and test patterns are propagated to and fed into each module. Test responses are similarly propagated to response analyzers placed only on primary outputs. For the propagation of test patterns and test responses, paths existing in the data path are utilized.\\
\indent Gosh es al. [88] introduce a novel scheme for testing RTL controller/data paths using BIST. The scheme uses the controller netlist and the data path of a circuit to extract a test control/data flow (TCDF) graph. This TCDF is used to derive a set of symbolic justification and propagation paths (known as test environment) to test some of the operations and variables present in it.
\subsection{Miscellaneous}
\indent Yogi et al. demonstrate in [89, 90] a new method of RTL test generation using spectral techniques. Test vectors generated for RTL faults are analyzed using Hadamard matrices to extract important features and new vectors are generated retaining those features.\\
\indent Karunaratne et al. [91]  take the advantage of the vectorized data paths in structured VLSI circuits and moves data in vector form. The search consists of a backward transfer process which resembles a vectorized consistency drive of D-algorithm over an iterative network in which cell copies are analogous to clock periods in sequential circuits. Network elements are RTL logic functions rather than individual gates and data lines may carry binary values as well as vectors.
\section{The GIF-PO RTL fault model}
\subsection{A simple example}

The proposed RTL model is best explained by using an example. Fig. 1 shows a simple circuit C1 with x = (a \& b) \textasciicircum\space c. C2 is functional equivalent to C1 (or permissible function) with x' = (a \& b \& !c) \textbar\space (!(a \& b) \& c). C2 can be an alternative representation, when nets like e = (a \& b \& !c) or f = (!(a \& b) \& c) can be shared with other functions in the digital circuit.\\
\indent Each gate in C1 has now a gate specific set of faults defined. These gate related faults are based on the individual input to output paths of each gate. The fault definition is given later.\\
\indent The proposed gate specific fault model makes it possible, that - based on the given function of C1 - a test set can be computed, which covers all stuck-at faults of all nets in any permissible function of C1, as for instance C2.

\begin{figure}[!t]
	\centering
	\includegraphics[width=3in]{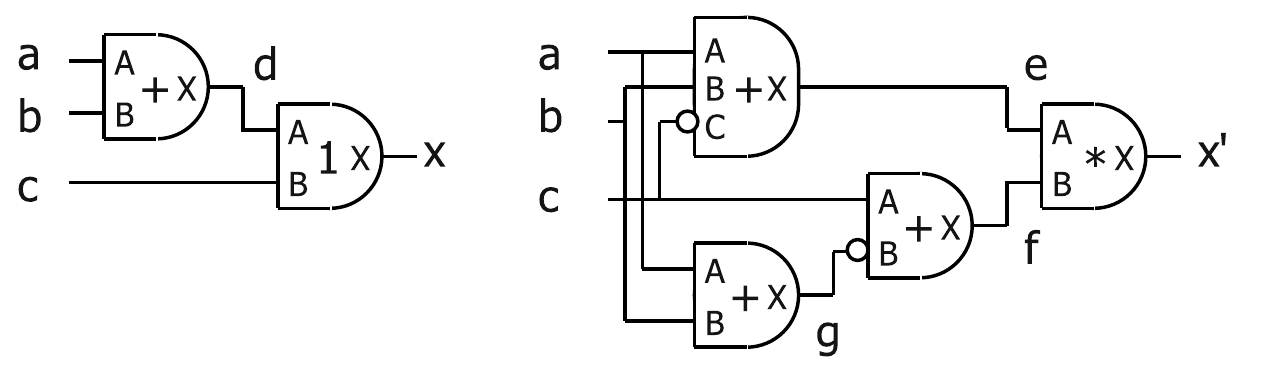}
	\caption{Circuit C1 and permissible circuit C2.}
	\label{fig1_c1c2}
\end{figure}

\begin{table}[!t]
	\setlength{\tabcolsep}{5pt}
	\renewcommand{\arraystretch}{1.3}
	\caption{Complex Gates in VHDL/Verilog}
	\label{table_example}
	\centering
	\begin{tabular}{l | l | l}
		Complex Gates & VHDL & Verilog \\ \hline \hline
		comb & not, and, or, xor & !, \&, \textbar, \textasciicircum, ...\\
		unary & & \&a, \textbar a, \textasciicircum a, ...\\
		mux & a(i) & a[i]\\
		demux & a(i) \textless= & a[i] \textless =\\
		if & it then else & ...?...:...; if else\\
		case & case (sel) when ...& case (sel) ...\\
		compare & /=, =, \textless, \textgreater, ... & !=, =, \textless, \textgreater, ...\\
		math & a + b, -, *, ... & a + b, -, *, ...\\
		shift & shl, shr & \textless\textless, \textgreater\textgreater\\
	\end{tabular}
\end{table}

\begin{table}[!t]
	\setlength{\tabcolsep}{5pt}
	\renewcommand{\arraystretch}{1.3}
	\caption{Example continued I}
	\label{table_example}
	\centering
	\begin{tabular}{l l l l | l l l l}
		\multicolumn{4}{c}{Y = Y AND B} & \multicolumn{4}{c}{Y = Y XOR B} \\ \hline \hline
		A & B & Y & GIF & A & B & Y & GIF \\ \hline \hline
		0 & 0 & 0 &     & 0 & 0 & 0 & A1, B1\\
		0 & 1 & 0 & A1  & 0 & 1 & 1 & A2, B2\\
		1 & 0 & 0 & B1  & 1 & 0 & 1 & A3, B3\\
		1 & 1 & 1 & A2, B2 & 1 & 1 & 0 & A4, B4\\
	\end{tabular}
\end{table}

\begin{table*}[t]
	\renewcommand{\arraystretch}{1.3}
	\caption{Example continued II}
	\label{table_example}
	\centering
	\begin{tabular}{l l l | l l | l | l | l l l | l l l l l}
		\multicolumn{3}{c}{} & \multicolumn{2}{c}{Circuit C1} & \multicolumn{2}{c}{} & \multicolumn{3}{c}{Circuit C1} & \multicolumn{5}{c}{Circuit C2} \\ \hline \hline
		a & b & c & AND    & XOR    & Ti  & Tii  & d & x & net faults Ti           & e & f & g & x' & net faults Ti \\ \hline \hline
		0 & 0 & 0 &        & A1, B1 &   &   & 0 & 0 &                         & 0 & 0 & 0 & 0  & \\
		0 & 0 & 1 &        & A2, B2 &   &   & 0 & 1 &                         & 0 & 0 & 1 & 1  & \\
		0 & 1 & 0 & A1	   & A1, B1 & 1 & 1 & 0 & 0 & a-1, c-1, d-1, x-1      & 0 & 0 & 0 & 0  & a-1, c-1, e-1, g-1, x'-1\\
		0 & 1 & 1 & A1     & A2, B2 &   & 2 & 0 & 1 &                         & 0 & 0 & 1 & 1  & \\
		1 & 0 & 0 & B1     & A1, B1 &   &   & 0 & 0 &                         & 0 & 0 & 0 & 0  & \\
		1 & 0 & 1 & B1     & A2, B2 & 2 &   & 0 & 1 & b-1, c-0, d-1, x-0      & 0 & 0 & 1 & 1  & b-1, c-0, f-1, g-0, x'-0\\
		1 & 1 & 0 & A2, B2 & A3, B3 & 3 & 3 & 1 & 1 & a-0, b-0, c-1, d-0, x-0 & 1 & 1 & 0 & 1  & a-0, b-0, c-1, e-0, x'-0\\
 		1 & 1 & 1 & A2, B2 & A4, B4 & 4 & 4 & 1 & 0 & a-0, b-0, c-0, d-0, x-1 & 0 & 1 & 0 & 0  & c-0, e-1, f-0, g-1, x'-1\\
	\end{tabular}
\end{table*}

\subsection{Using more complex gates}
\indent RTL languages are based on complex gates. Table I gives an overview of complex gates in VHDL or Verilog, which do generate logic during synthesis. Some can be composed out of less complex gates. A comparator can be build out of XOR, INV and AND gates. A complex adder gate can be replaced by a series of bit-wise adders, which themselves can be replaced by less complex gates, etc.. The proposed gate specific fault model can be applied directly on any given complex gate.\\
\subsection{The gate inherent fault (GIF)}
\indent The proposed RTL model is based on gate inherent faults (GIF), which are solely related to the different signal propagation paths through a gate. They are not related to specific stuck-at values of any input or output pins of the gate, and therfore not related to any net/signal in the RTL source code.\\
\indent When a path between a register (design input) and another register (design output) only consists out of a simple assignments, then this assignment has to be considered as a complex gate and GIF must be allocated for it.
\subsection{The GIF-GO model definition}
\indent Under the proposed GIF-GO model, a GIF is described by a quadruple(gi, go, i, $\alpha$) where gi is a gate input, go is a gate output, i is an index and $\alpha$ $\in$ \{0,1\}. The fault (gi, go, i, $\alpha$) is detected by a test t that satisfies the following conditions.\\
\begin{enumerate}[\IEEEsetlabelwidth{12)}]
	\item The test t detects the path fault gi to go with index i (gi-go-i).\\
	\item The fault free value of output go under t is $\alpha$.\\
	\item In the presence of the fault gi-go-i, the output value go = !$\alpha$.\\
\end{enumerate}
\indent In other words, t propagates the effects of a gi-go path fault with index i to the gate output go. The output's value is $\alpha$ in the fault free circuit and !$\alpha$ in the presence of the fault. 
\subsection{Logic duplication}
\indent An important element of RTL synthesis is logic duplication. Duplicated logic can generate net faults which are not detected when a test set is used that is based on the GIF-GO model. Therefore the final RTL fault model needs to consider logic duplication. A gate is an element of a network of combinatorial logic. All outputs of this combinatorial logic are called primary outputs (PO). In case of a sequential netlist, register data inputs are considered as PO as well.
\subsection{The GIF-PO model definition}
\indent Under the proposed GIF-PO model, a GIF is described by a quintuple(gi, go, i, j, $\alpha$) where gi is a gate input, go is a gate output, i is an index, j is a primary output and $\alpha$ $\in$ \{0,1\}. The fault (gi, go, i, j, $\alpha$) is detected by a test t that satisfies the following conditions.\\
\begin{enumerate}[\IEEEsetlabelwidth{12)}]
\item The test t detects the path fault gi to go with index i (gi-go-i).\\
\item The fault free value of primary output j under t is $\alpha$.\\
\item In the presence of the fault gi-go-i, the primary output value j = !$\alpha$.\\
\end{enumerate}
\indent In other words, t propagates the effects of a gi-go path fault with index i to the primary output j. The primary output's value is $\alpha$ in the fault free circuit and !$\alpha$ in the presence of the fault. A single test t can cover multiple GIF. 
\subsection{The example continued}
\indent The proposed GIF-PO model is now applied on the given example. Table II lists the GIF for an AND and an XOR gate (Fig. 1, C1). The GIF can be simply indexed (1, 2, ..., N), but in this case, the name of the input pin, to which the GIF is sensitive to, is added for better readability.\\
\indent Table III lists all possible combinations of the circuit’s inputs a, b and c. The ''Circuit C1'' columns show the GIF coverage for the AND and the XOR gate when the relevant input values are applied on C1. All GIF have a single primary output x (x' for C2 respectively).\\
\indent An algorithm now selects a minimum test set T that covers all GIF for both gates. Two different test sets (Ti, Tii) are found. The relevant cycles for each test set are numbered in the Ti (Tii) column. The ''Circuit C1'' column now shows, that all net faults of C1 are covered by the selected Ti. A net fault is denoted as net name and stuck-at value, separated by a dash. Ti can be reduced (by cycle 3 or 4), based on the given netlist C1. Ti achieves 100\% stuck-at coverage on C1 and C2, as the column ''Circuit C2'' shows. Same is true for Tii. Test sets can also be generated based on C2 and applied on C1 as well.

\subsection{The GIF-PO model on RTL}
\indent Digital designs written on RTL usually contain complex gates (Table I). The GIF for each complex gate are extracted and uniquified for each one of their individual primary outputs. Three more steps need to be done to identify the relevant GIF-PO of the design. Firstly, constants need to be (forward-) propagated through the RTL logic. Secondly, open pins need to be (backward-) propagated through the RTL logic.\\
\indent In a third step, unreachable GIF-PO must be identified. There are multiple ways to identify unreachable GIF-PO. One trivial way is based on the fact, that when all input permutations are tested, then all uncovered GIF-PO are redundant, as shown in the following multiplier example. Ohtake et al. propose in [92] a method for unsensitizable path identification using high level design information. The technique shown in [92] can be used to identify unreachable GIF-PO. This paper proposes in the next Section a semi-automatic test pattern generation flow, which allows the user to mark unreachable GIF-PO manually. In future work, a fully automated RTL TPG based on the GIF-PO model will be shown, which automatically identifies unreachable GIF-PO using modern formal and functional verification techniques. 

\subsection{An adder example}
\begin{table}[!t]
	\setlength{\tabcolsep}{5pt}
	\renewcommand{\arraystretch}{1.3}
	\caption{GIF-PO of a Half Adder}
	\label{table_example}
	\centering
	\begin{tabular}{l l | l l l | l l l}
		A & B & S & GIF-PO(S)  & ID & CO & GIF-PO(CO) & ID \\ \hline \hline
		0 & 0 & 0 & A1, B1 & 1   & 0  &        & \\
		0 & 1 & 1 & A1, B2 & 2   & 0  & A5     & 1 \\
		1 & 0 & 1 & A3, B3 & 3   & 0  & B5     & 2\\
		1 & 1 & 0 & A4, B4 & 4   & 1  & A6, B6 & 3\\
	\end{tabular}
\end{table}

\begin{table}[!t]
	\setlength{\tabcolsep}{5pt}
	\renewcommand{\arraystretch}{1.3}
	\caption{GIF-PO of a 1-Bit Full Adder}
	\label{table_example}
	\centering
	\begin{tabular}{l l l | l l l | l l l}
		CI & A & B & S & GIF-PO(S)  & ID & CO & GIF-PO(CO) & ID \\ \hline \hline
		0 & 0 & 0 & 0 & A1, B1, C1 & 1   & 0  &          & \\
		0 & 0 & 1 & 1 & A1, B2, C2 & 2   & 0  & A9, C9   & 1 \\
		0 & 1 & 0 & 1 & A3, B3, C3 & 3   & 0  & B9, C10  & 2 \\
		0 & 1 & 1 & 0 & A4, B4, C4 & 4   & 1  & A10, B10 & 3 \\
		1 & 0 & 0 & 0 & A5, B5, C5 & 5   & 0  & A11, B11 & 4 \\
		1 & 0 & 1 & 1 & A6, B6, C6 & 6   & 1  & B12, C11 & 5 \\
		1 & 1 & 0 & 1 & A7, B7, C7 & 7   & 1  & A12, C12 & 6 \\
		1 & 1 & 1 & 0 & A8, B8, C8 & 8   & 1  &          & \\
	\end{tabular}
\end{table}

\indent Complex gates can be split into sub gates. For this N-bit wide adder (N-ADD) example, we apply the GIF-PO model on a half adder (HA) and on a 1-bit wide full-adder (FA). Table IV (Table V) shows the GIF-PO of a HA (FA), assuming that S and CO are PO of the HA (FA). The GIF can alternatively be grouped and indexed for each PO as shown.\\
\indent The HA and multiple FA are combined to form the N-ADD. The resulting PO of the N-ADD are the S(0, …, N-1) ports. The GIF related to the CO outputs of the HA (FA) must be duplicated for each of the PO, which depends on the internal CO signals. The CO value of the HA is propagated to the S(1, …, N-1) ports. The CO value of the first FA (bit 1) is propagated to the S(2, …, N-1) ports and so on. The related GIF are duplicated accordingly.\\
\indent A 64-bit adder has 12415 GIF-PO. A simple mechanism can be applied to generate a test set with 100\% GIF-PO coverage. It takes only 193 functional cycles to achieve 100\% GIF-PO coverage and 100\% gate level stuck-at coverage of a 64-ADD, no matter how it is synthesized. These pattern can be further reduced when a concrete implementation is available.
\subsection{A multiplier example}
\begin{figure}[htb]
	\begin{center}
		\includegraphics[width=0.4\textwidth]{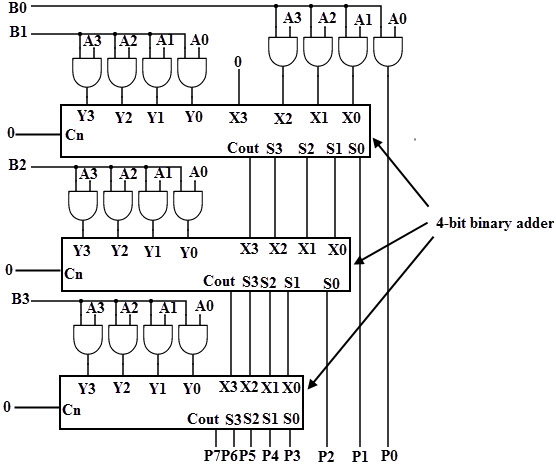}
	\end{center}
	\caption{Schematic view of a 4-bit wide multiplier using AND gates and three 4-bit wide adders.}
	
	\label{fig1_c1c2}
\end{figure}

\indent A multiplier is used to elaborate on untestable GIF-PO and untestable stuck-at coverage points. The multiplier can be seen as a complex gate and the relevant GIF-PO can be defined for it. Alternatively, we generate the 8-bit wide multiplier (N-MUL) out of AND gates and seven (N-1) 8-ADD, whereas each 8-ADD is based on one HA and seven (N-1) FA.This RTL circuit has 1935 GIF-PO and builds a directed graph which holds therefore the potential of untestable GIF-PO. A schematic view of a fragmented 4-bit wide multiplier is shown in Fig 2.\\
\indent After applying all input permutations, only 1886 GIF-PO are covered, the rest are considered as unreachable. Only 906 functional cycles actually contribute to this GIF-PO coverage (without ordering or optimization). They define one possible test set T to achieve 100\% stuck-at coverage.\\
\indent An initial gate level representation of the 8-MUL is built out of AND gates and seven 8-ADD again. This time, the 8-ADD are generated based on AND, OR and INV gates, whereas each carry-in for each individual adder bit is generated based on its very own individual AND-OR tree. This complex AND-OR representation of an 8-MUL netlist has 3264 stuck-at coverage points. After applying the test pattern generated on RTL based on the GIF-PO model (T), only 3236 faults are covered. The uncovered faults are considered as untestable and the relevant logic can be seen as redundant.\\
\indent In the final step, all nets with uncovered faults are either set to constant 0 or 1, depending on their actual coverage result. This new netlist is then used to simulate all possible input permutations. The modified circuit works correctly and only 521 test cycles (out of the 906 functional cycles of T) are needed to achieve 100\% stuck-at coverage on the gate level netlist.
\subsection{Difference to related RTL ATPG models}
\indent One related RTL fault model is the PI-PO fault model, shown by Pomeranz et al. in [39]. The PI-PO fault model is related to the primary inputs and primary outputs of a given logic block. The Input/Output TRansition (TRIO) model shown by Kang et al. in [40] proposes similar primary input to primary output paths and toggling considerations to estimate the gate level coverage based on RTL. In contrast to both models, the GIF-PO model considers the gate input and gate output of each gate (and therefore its gate inherent logic path) and combines it with the relevant primary outputs of the given combinatorial network.\\
\indent It is questionable, if test sets for all path sensitivities between primary inputs and primary outputs of a complex netlist can be found within a reasonable time. If so, ATPG would not be a problem at all. The author of this paper therefore assumes, that in best case, the PI-PO and TRIO fault models can only be used to estimate the fault coverage based on a given pattern set, as explicitly mentioned in [39, 40].\\
\indent The proposed GIF-PO model allows the calculation of the maximum GIF-PO number of a given logic directly, by using a specific number for each individual complex gate and the number of primary outputs, connected to the individual gate output. The goal of the GIF-PO model is to generate test pattern which achieve a 100\% GIF-PO coverage.\\
\indent Another related RTL fault model is described in the work of Patel et al. in [36]. Their approach involves the generation of a two-level AND-OR, or OR-AND representation from a circuit’s functional description, which then serves as an abstract model for the generation of test vectors. The work by Jacob et al. [37] further develops the two-level AND-OR based matrix model for its usage on FSMs. The disadvantage of the AND-OR usage is, that the gate-level test generation can be particularly expensive due to large reconvergent fanouts. The similarity to the GIF-PO model is, that the fault coverage relies solely on the functionality of gates and that it is not related to the individual nets of a given circuit representation. The main difference is, that the GIF-PO model targets a network of (complex) gates without a conversion to AND-OR logic and considers primary outputs (and therefore logic duplication).\\
\indent The empirical observations is, that a test set with 100\% GIF-PO based coverage achieves 100\% stuck-at fault coverage on any permissible netlist. This observation is unique and has not been reported before by any other alternative RTL ATPG model.

\section{A semi-ATPG framework}

\begin{figure}[!t]
	\centering
	\includegraphics[width=2in]{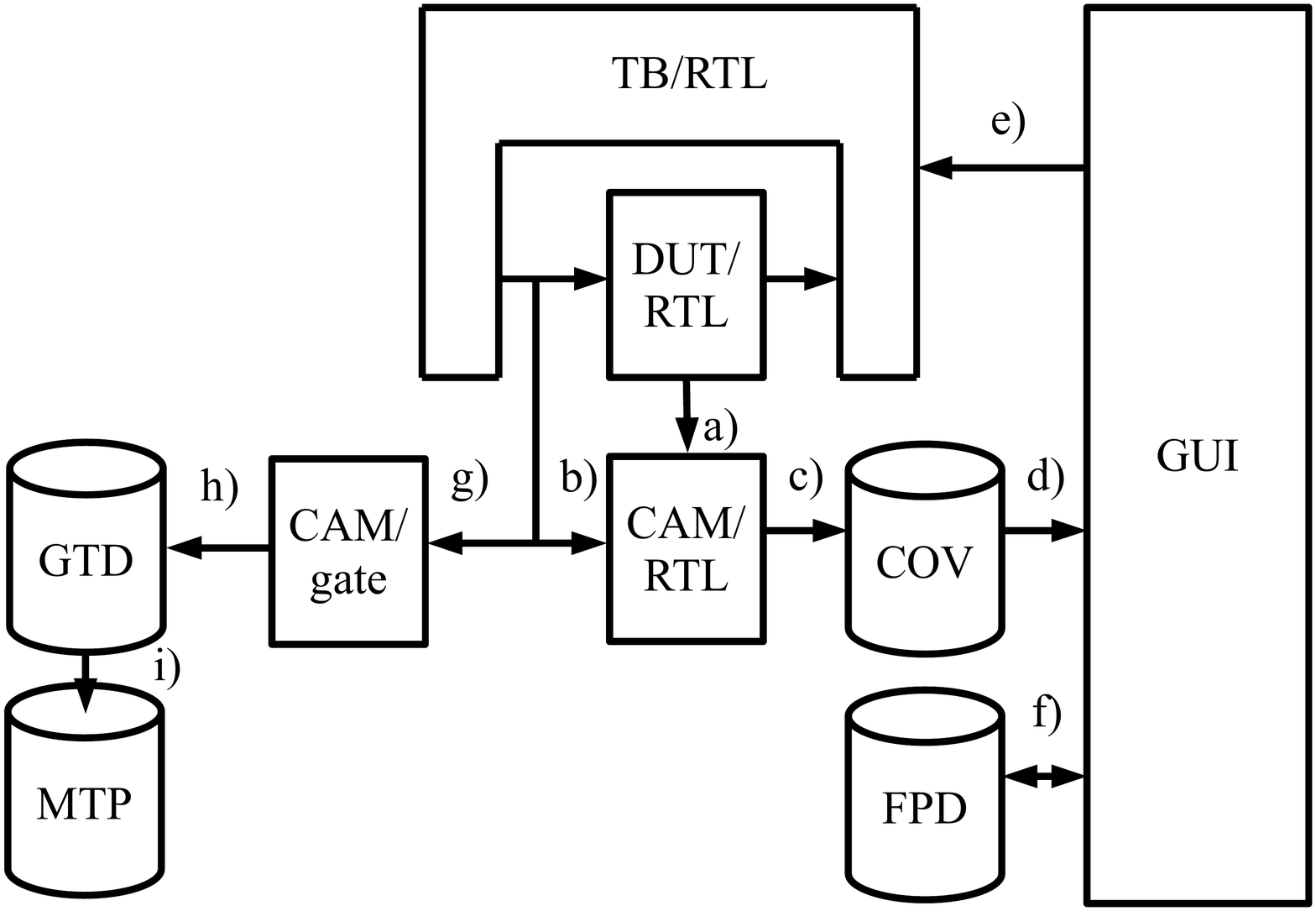}
	\caption{Overview of the used semi-ATPG framework.}
	\label{fig2_framework}
\end{figure}

\subsection{Introduction of the used semi-ATPG framework}
\indent Fig. 3 gives an overview of the used flow with its individual steps (a, b, ...). The design under test (DUT) written on RTL is controlled by an RTL testbench (TB) for simulation. In a first step (a), the DUT is converted into a cycle accurate model (CAM/RTL). The author shows in [93], how such a model can be generated. The RTL-CAM is enhanced to define and propagate GIF-PO during simulation.\\
\indent At the beginning of each simulation run, constant values are propagated through the design. Unconnected (open) signals are backward propagated through the design. The CAM/RTL is stimulated (b) with the same input pattern as the DUT and dumps (c) the GIF-PO coverage into a coverage database (COV). The GIF-PO coverage and the relevant RTL source code can be made visible (d) by using a GUI. This information is used by the designer / verificationist to define false paths and unreachable GIF-PO. It is also used to improve (e) the testcases in order to achieve 100\% GIF-PO coverage. The user defined redundant GIF-PO are stored in the false path database (FPD) to be reused in the next run (f).\\
\indent A second CAM is generated based on the gate level netlist (CAM/gate), which is then stimulated (g) with the same stimuli as the DUT. During this gate level simulation, a database is build (h), which contains the general test set database (GTD) and all necessary information to generate relevant test pattern on the implemented test structure. This database is then optimized (i) to generate a minimal test pattern (MTP) for the final production test.

\begin{table*}[t]
	\setlength{\tabcolsep}{5pt}
	\renewcommand{\arraystretch}{1.3}
	\caption{Coverage Results}
	\label{table_example}
	\centering
	\begin{tabular}{l l l l l l l l l}
		& GIF-PO & GIF-PO & Functional & Pattern & Coverage & Nets  & Pattern& Coverage\\ 
		& & redundant & cycles & RTL & GIF-PO & & netlist & stuck-at	 \\ \hline \hline
		64-add (AO)   & 12415  & 0                & 445               & 193           & 100\%           & 21054 & 133               & 100\%\\
		8x8-mult (AO) & 1935   & 49               & 65535             & 906           & 100\%           & 3236  & 521               & 100\%\\
		b01           & 368    & 106              & 40                & 34            & 100\%           & 56    & 15                & 100\%\\
		b02           & 184    & 52               & 14                & 12            & 100\%           & 33    & 10                & 100\%\\
		b06           & 573    & 172              & 32                & 24            & 100\%           & 67    & 14                & 100\%\\
	\end{tabular}
\end{table*}

\subsection{Discussion on the proposed flow}
\indent Existing code coverage tools can be enhanced to support the proposed flow. It guides the user to redundant RTL code, when GIF-PO coverage points cannot be covered. It helps to discover redundant logic in the netlist, when 100\% stuck-at coverage of the netlist cannot be reached, although a test set with 100\% GIF-PO coverage is applied.\\
\indent With this flow, a test set database can be generated before the design is synthesized. The RTL test sets can also be provided as a deliverable for an IP. The gate level test set can be optimized based on the concrete implementation. When a random pattern generator is used for the initial test sequence, the database can help to generate a test set for stuck-at faults, which are hard to cover.
\subsection{Difference to related work}
\indent Known code coverage tools in the EDA domain also concentrate on complex gates. To the best of the author\textquoteright s knowledge, non of them uses the GIF-PO model. They derive for instance statement, branch, path and expression based coverage points. It is not defined, that they relate the code coverage points to primary outputs (PO), nor require that a fault must be propagated to a PO to be considered as covered.\\
\indent Ferrandi et al. show in [94] functional test generation for behavioral sequential models. They use statement, branch, condition and path coverage to achieve high functional coverage and tweak it to get optimized experimental results for multi-cycle gate level coverage.\\
\indent To a certain extent, similar flows are also proposed in the already mentioned work as well. The differences are, that [2] uses event and hit counts, and [3] output derivations as the individual coverage matrices. Non of them uses the GIF-PO model. [7] uses RTL stuck-at fault injection and [10] uses error injection, whereas the proposed flow in this paper works without an injection based technique. [5] concentrates on dark corners of the design, this paper targets 100\% GIF-PO coverage of the complete design.

\section{Results}

\begin{figure}[!t]
	\centering
	\includegraphics[width=3in]{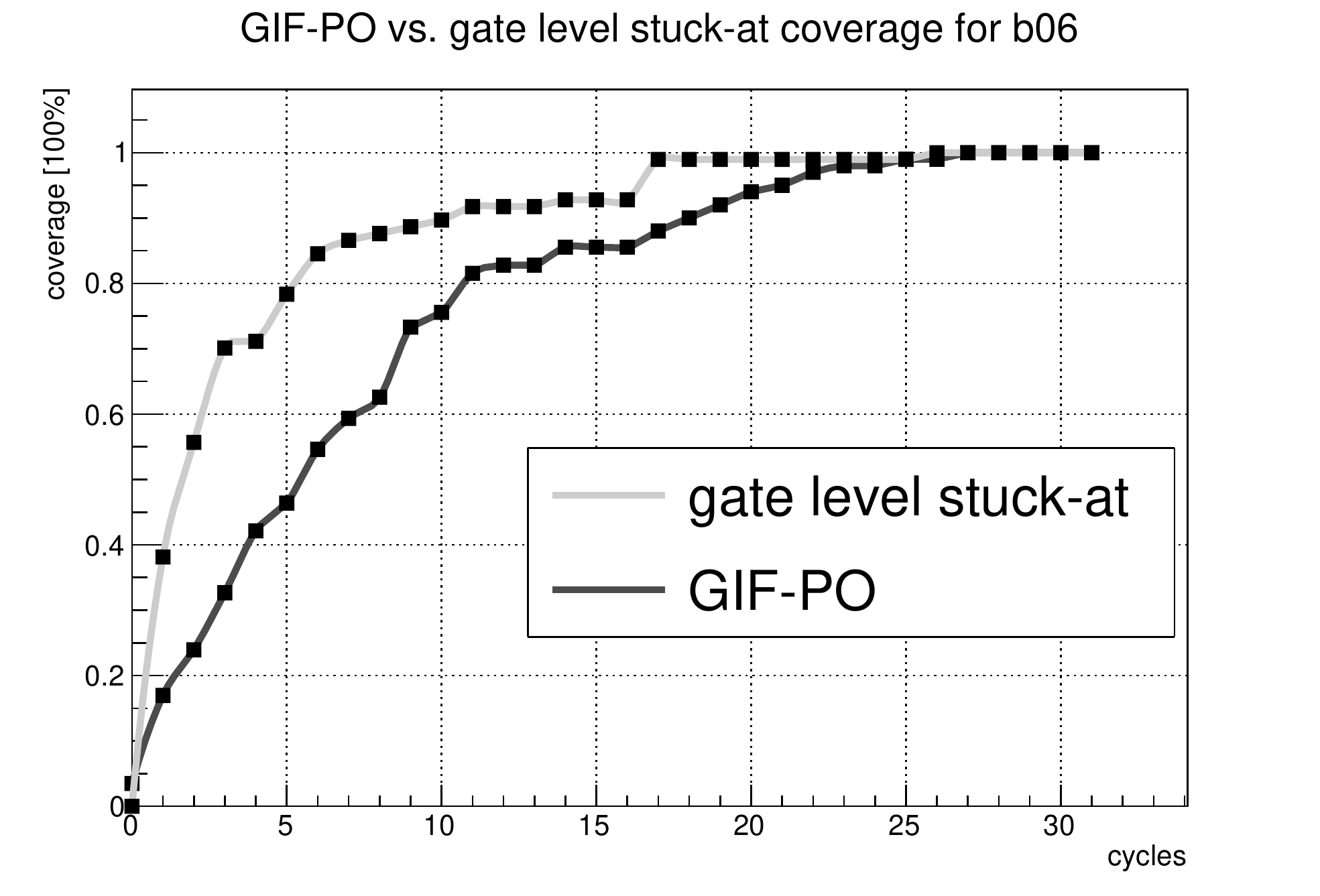}
	\caption{ GIF-PO and stuck-at correlation during functional simulation for b06.}
	\label{fig2_framework}
\end{figure}

\indent The results are based on the following testcases. A 64-bit wide adder is built out of a simple ''+'' construct on RTL and the related netlist is based on AND, OR and INV gates. An 8-bit wide multiplier netlist is generated based on AND gates and fifteen 16-bit adders, which themselves are made out of AND, OR and INV gates. Redundant logic is removed similar to the mechanism shown in Section III.J. The b01, b02 and b06 circuits with their standard netlists of the ITC\textquoteright 99 testcases are used.\\
\indent Table VI shows the GIF-PO coverage points for each testcase and the redundant GIF-PO, which are removed during the initial constant/open signal propagation and false path detection phase.\\
\indent in the next column the amount of functional cycles to achieve 100\% GIF-PO coverage in simulation is listed. The column ''Pattern (RTL)'' lists the number of functional cycles (T), which add GIF-PO coverage during simulation (without ordering). All testcases achieve 100\% GIF-PO coverage.\\
\indent Next, the number of nets of the provided netlist is listed. The column ''Pattern (netlist)'' shows the number of cycles which add stuck-at coverage while simulating T on gate level. All testcases achieve 100\% stuck-at coverage. \\
\indent Fig. 4 shows the GIF-PO and stuck-at correlation during testbench driven functional simulation of b06. The flat sections in both gate level stuck-at and GIF-PO curves show, that pattern can be removed when converted into test pattern.
\section{Concluding Remarks}
\indent How the proposed model based on inherent faults of (complex) gates can be combined with classical ATPG techniques to generate a test set for stuck-at fault testing of nets  needs further discussion. This paper proposes a semi-automatic pattern generation process based on functional tests to lay the foundation for future work on generating sequential test sets. It remains to be shown, how the GIF-PO model can be incorporated into more advanced techniques in the field of automatic functional pattern generation (AFPG).\\
\indent How register mapping (for instance for one-hot FSM), register duplication and balancing impacts the proposed RTL ATPG model, and how sequential test sets help to overcome potential limitations also needs to be discussed. In the context of generating 100\% stuck-at fault coverage based on advanced AFPG techniques, aspects like transition faults, realistic path selection for at-speed testing, a reduced overtesting and a reduced false path testing will need to be evaluated. The proposed GIF-PO ATPG model can also be used for BIST implementations, which are optimized on RTL.

\ifCLASSOPTIONcaptionsoff
  \newpage
\fi




\begin{thebibliography}{95}

\bibitem{IEEEhowto:1}
I. Harris, ''Hardware-Software Covalidation: Fault Models and Test Generation'', Proc. Sixth IEEE Intern. High-Level Design Validation and Test Workshop, 7-9 Nov. 2001, Monterey, CA, USA, pp. 151-156.
\bibitem{IEEEhowto:2}
S. Park, L. Chen, P. Parvathala, S. Patil, and I. Pomeranz, ''A Functional Coverage Metric for Estimating the Gate-Level Fault Coverage of Functional Tests'', IEEE Intern. Test Conf., ITC '06, Oct. 2006, Santa Clara, CA, USA, pp. 1-10.
\bibitem{IEEEhowto:3}
H. Fang, K. Chakrabarty, A. Jas, S. Patil, C. Tirumurti, ''RT-Level Deviation-Based Grading of Functional Test Sequences'', 27th IEEE VLSI Test Symposium, VTS '09, 3-7 May 2009, Santa Cruz, CA, USA, pp. 264-269.
\bibitem{IEEEhowto:4}
H. Fang, K. Chakrabarty, A. Jas, S. Patil, C. Tirumurti, ''RT-Level Deviation-Based Grading of Functional Test Sequences'', IEEE Trans. on VLSI, vol. 20, no. 10, October 2012, pp. 1890-1894.
\bibitem{IEEEhowto:5}
C. Hobeika, C. Thibeault, and J.-F. Boland, ''Use of Structural Tests in RTL Verification'', IEEE 1st Microsystems and Nanoelectronics Research Conf., MNRC 2008, 15 Oct. 2008, Ottawa, Ont., Canada, pp. 133-136.
\bibitem{IEEEhowto:6}
A. Sanyal, K. Chabrabarty, M. Yilmaz, and H. Fujiwara, ''RT-Level Design-for-Testability and Expansion of Functional Test Sequences for Enhanced Defect Coverage'', IEEE Intern. Test Conf. ITC, 2-4 Nov. 2010, Austin, TX, USA, pp. 1-10.
\bibitem{IEEEhowto:7}
W. Mao, and R. Gulati, ''Improving Gate Level Fault Coverage by RTL Fault Grading'', Proc. of Intern. Test Conf., ITC 1996, 20-25 Oct. 1996, Washington, DC, USA, pp. 150-159.
\bibitem{IEEEhowto:8}
M. Santos, F. Goncalves, I. Teixeira, and J Teixeira, ''RTL-based Functional Test Generation for High Defects Coverage in Digital SOCs'', Proc. IEEE European Test Workshop, 23-26 May 2000, Cascais, Portugal, pp. 99-104.
\bibitem{IEEEhowto:9}
Y. Zhang, H. Li, and X. Li, ''Software-Based Self-Testing of Processors Using Expanded Instructions'', IEEE 19th Asian Test Symposium, ATS ‘10, 1-4 Dec. 2010, Shanghai, China, pp. 415-420.
\bibitem{IEEEhowto:10}
F. Ferrandi, G. Ferrara, G. Fornara, F. Fummi, and D. Sciuto, ''Testability Alternatives Exploration through Functional Testing'', Proc. of 18th IEEE VLSI Test Symposium, 30 Apr. - 4 May 2000, Montral, Que. Canada, pp. 423-428. 
\bibitem{IEEEhowto:11}
V. Chaiyakul, D. Gajski, and L. Ramachandran, ''High-Level Transformations for Minimizing Syntactic Variances'', 30th Conf. on Design Automation, 14-18 June 1993, Dallas, TX, USA, pp. 413-418
\bibitem{IEEEhowto:12}
I. Gosh, and M. Fujita, ''Automatic Test Pattern Generation for Functional Register-Transfer Level Circuits Using Assignment Decision Diagrams'', IEEE Trans. on CAD, vol. 20, no. 3, March 2001, pp. 402-415.
\bibitem{IEEEhowto:13}
I. Gosh, and M. Fujita, ''Automatic Test Pattern Generation for Functional Register-Transfer Level Circuits Using Assignment Decision Diagrams'', Proc. Of DAC, 5-9 June 2000, LA, CA, USA, pp. 43-48.
\bibitem{IEEEhowto:14}
L. Zhang, I. Ghosh, and M. Hsiao, ''Efficient Sequential ATPG for Functional RTL Circuits'', Proc. of Intern. Test Conf.  ITC, 30th Sept.-2nd Oct. 2003, pp. 290-298.
\bibitem{IEEEhowto:15}
L. Zhang, M. Hsiao, and I. Ghosh, ''Automatic Design Validation Framework for HDL Description via RTL ATPG'', Proc. of 12th Asian Test Symposium, 16-19 Nov. 2003, Xi'an, China, pp. 148-153.
\bibitem{IEEEhowto:16}
L. Zhang, I. Ghosh, and M. Hsiao, ''A Framework for Automatic Design Validation of RTL Circuits Using ATPG and Observability-Enhanced Tag Coverage'', IEEE Trans. on CAD, vol. 25, no. 11, Nov. 2006, pp. 2526-2538.
\bibitem{IEEEhowto:17}
L. Lingappan, S. Ravi, N. Jha, ''Test Generation for Non-separable RTL Controller-datapath Circuits using a Satisfiability based Approach'', Proc. of 21st Intern. Conf. on Computer Design ICCD, 13-15 Oct. 2003, San Jose, CA, USA, pp. 187-193.
\bibitem{IEEEhowto:18}
L. Lingappan, S. Ravi, N. Jha, ''Satisfiability-Based Test Generation for Nonseparable RTL Controller-Datapath Circuits'', IEEE Trans. on CAD, vol. 25, no. 3, March 2006, pp. 544-557.
\bibitem{IEEEhowto:19}
L. Lingappan, and N. Jha, ''Satisfiability-Based Test Program Generation and Design for Testability for Microprocessors'', IEEE Trans. on VLSI, vol. 15, no. 5, May 2007, pp. 518-530.
\bibitem{IEEEhowto:20}
M. Obien, S. Ohtake, and H. Fujiwara, ''Constrained ATPG for Functional RTL Circuits Using F-Scan'', IEEE Intern. Test Conference ITC 2010, 2-4 Nov. 2010, Austin, TX, USA, pp. 1-10.
\bibitem{IEEEhowto:21}
L. Lingappan, V. Gangaram, N. Jha, and S. Chakrvarty, ''Fast Enhancement of Validation Test Sets to Improve Stuck-at Fault Coverage for RTL circuits'', 20th Intern. Conf. on VLSI Design VLSID'07, 6-10 Jan. 2007, Bangalore, India, pp. 504-512.
\bibitem{IEEEhowto:22}
L. Lingappan, V. Gangaram, N. Jha, and S. Chakrvarty, ''Fast Enhancement of Validation Test Sets for Improving the Stuck-at Fault Coverage of RTL Circuits'', IEEE Trans. on VLSI, vol. 17, no. 5, 2009, pp. 697-708.
\bibitem{IEEEhowto:23}
L. Lingappan, and N. Jha, ''Unsatisfiability based Efficient Design for Testability Solution for Register-transfer Level Circuits'', Prod. of 23rd IEEE VLSI Test Symposium VTS'05, 1-5 May 2005, Palm Springs, CA, USA, pp. 418-423.
\bibitem{IEEEhowto:24}
L. Lingappan, and N. Jha, ''Efficient Design for Testability Solution Based on Unsatisfiability for Register-transfer Level Circuits'', IEEE Trans. on CAD, vol. 26, no. 7, July 2007, pp. 1339-1345.
\bibitem{IEEEhowto:25}
C.-H. Chen, and D. Saab, ''A Novel Behavioral Testability Measure'', IEEE Trans. On CAD, vol. 12, no. 12, December 1993, pp. 1960-1970.
\bibitem{IEEEhowto:26}
I. Gosh, A. Raghunathan, and N. Jha, ''A Design for Testability Technique for RTL Circuits Using Control/Data Flow Extraction'', Intern. Conf. on CAD, ICCAD’96, 10-14 Nov. 1996, San Jose, CA, USA, pp. 329-336.
\bibitem{IEEEhowto:27}
I. Gosh, A. Raghunathan, and N. Jha, ''A Design for Testability Technique for RTL Circuits Using Control/Data Flow Extraction'', IEEE Trans. on CAD, vol. 17, no. 8, August 1998, pp. 706-723.
\bibitem{IEEEhowto:28}
J. Lee, and J. Patel, ''Architectural Level Test Generation for Microprocessors'', IEEE Trans. on CAD, vol. 13, no. 10, October 1994, pp. 1288-1300.
\bibitem{IEEEhowto:29}
J. Raik, and R. Ubar, ''Fast Test Pattern Generation for Sequential Cirtuits Using Decision Diagram Representations'', Journal of Electronic Testing, Kluwer Academic Publishers, 2000, pp. 213-226.
\bibitem{IEEEhowto:30}
J. Raik, and R. Ubar, ''Sequential Cirtuit Test Generation Using Decision Diagram Representations'', Proc. Design, Automation and Test in Europe Conf., DATE 1999, 9-12 March 1999, Munich, Germany, pp. 736-740.
\bibitem{IEEEhowto:31}
F. Corno, G. Cumani, M. Reorda, and G. Squillero, ''Effective Techniques for High-level ATPG'', Proc. of 10th Asian Test Symposium, ATS’01, Kyoto, Japan, pp. 225-230.
\bibitem{IEEEhowto:32}
H. Wada, T. Masuzawa, K. Saluja, and H. Fujiwara, ''Design for Strong Testability of RTL Data Paths to Provide Complete Fault Efficiency'', 13th Intern. Conf. on VLSI, 3-7 Jan. 2000, Calcutta, India, pp. 300-305.
\bibitem{IEEEhowto:33}
F. Fallah, S. Devadas, and K. Keutzer, ''Functional Vector Generation for HDL Models Using Linear Programming and Boolean Satisfiability'', IEEE Trans. CAD, vol. 20, no. 8, August 2001, pp. 994-1002.
\bibitem{IEEEhowto:34}
Z. Yin, Y. Min, and X. Li, ''An Approach to RTL Fault Extraction and Test Generation'', 10th Asian Test Symposium, 19-21 Nov. 2001, Kyoto, Japan, pp. 219-224.
\bibitem{IEEEhowto:35}
S. Yadavalli, I. Pomeranz, and S. Reddy, ''MUSTC-Testing: Multi-Stage-Combinational Test Scheduling at the Register-Transfer-Level'', Proc. of 8th Intern. Conf. onVLSI Design 1995, 4-7 Jan. 1995, New Delhi, India, pp. 110-115.
\bibitem{IEEEhowto:36}
U. Dave, and J. Patel, ''A Functional-Level Test Generation Methodology Using Two-Level Repre-sentations'', 26th Conf. on Design Automation, DAC, 25-29 June 1989, Las Vegas, NV, USA, pp. 722-725.
\bibitem{IEEEhowto:37}
J. Jacob, and V. Agrawal, ''Functional Test Generation for Sequential Circuits'', The Fifth Intern. Conf. on VLSI Design,  4-7 Jan. 1992, Bangalore, India, pp. 17-24.
\bibitem{IEEEhowto:38}
F. Corno, G. Cumani, M. Reorda, and G. Squillero, ''An RT-level Fault Model with High Gate Level Correlation'', Proc. IEEE Intern. High-Level Design Validation and Test Workshop, 8-10 Nov. 2000, Berkeley, CA, USA, pp. 3-8.
\bibitem{IEEEhowto:39}
I. Pomeranz, and S. Reddy, ''Estimating the Relative Single Stuck-at Fault Coverage of Test Sets for a Combinational Logic Block from its Functional Description'', Proc. Sixth IEEE Intern. High-level Design Validation and Test Workshop, HLDVT’01, 7-9 November 2001, pp. 31-35.
\bibitem{IEEEhowto:40}
J. Kang, S. Seth, and V. Gangaram, ''Efficient RTL Coverage Metric for Functional Test Selection'', 25th IEEE VLSI Test Symposium, 6-10 May 2007, Berkeley, CA, USA, pp. 318-324.
\bibitem{IEEEhowto:41}
G. Jervan, Z. Peng, O. Goloubeva, M. Reorda, and M. Violante, ''High-Level and Hierarchical Test Sequence Generation'', Seventh IEEE Inter. High-Level Design Validation and Test Workshop, 27-29 Oct. 2002, Cannes, France, pp. 169-174.
\bibitem{IEEEhowto:42}
T. Sarfert, R. Markgraf, E. Trischler, and M. Schulz, ''Hierarchical Test Pattern Generation Based on High-Level Primitives'', Proc. Intern. Test Conf. 1989, 29-31 Aug. 1989, Washington, DC, USA, pp. 470-479.
\bibitem{IEEEhowto:43}
T. Sarfert, R. Markgraf, M. Schulz, and E. Trischler, ''Hierarchical Test Pattern Generation Based on High-Level Primitives'', IEEE Trans. on CAD, vol. 11, issue 1, Jan. 1992, pp. 34-44.
\bibitem{IEEEhowto:44}
V. Vedula, and J. Abraham, ''A Novel Methodology for Hierarchical Test Generation using Functional Constraint Composition'', IEEE Intern. High-Level Design Validation and Test Workshop, 8-10 Nov. 2000, Berkeley, CA, USA, pp. 9-14.
\bibitem{IEEEhowto:45}
J.Lee, and J. Patel, ''Hierarchical Test Generation Under Architectural Level Functional Constraints'', IEEE Trans. CAD, vol. 15, no. 9, September 1996, pp. 1144-1151.
\bibitem{IEEEhowto:46}
Y. Makris, and A. Orailoglu, ''Property-Based Testability Analysis for Hierarchical RTL Designs'', Proc. of 6th IEEE Intern. Conf. on Electronics, Circuits and Systems, ICECS’99, 5-8 September 1999, Pafos, Cyprus, pp. 1089-1092
\bibitem{IEEEhowto:47}
Y. Makris, and A. Orailoglu, ''DFT Guidance through RTL Test Justification and Propagation Analysis'', Proc. Intern. Test Conf. 1998, 18-23 October 1998, Washington, DC, USA, pp. 668-677.
\bibitem{IEEEhowto:48}
Y. Makris, J. Collins, A. Orailoglu, and P. Vishakantaiah, ''TRANSPARENT: A System for RTL Testability Analysis, DFT Guidance and Hierarchical Test Generation'', Proc. of the IEEE Custom Integrated Circuits Conf. 1999, 16-19 May 1999, San Diego, CA, USA, pp. 159-162
\bibitem{IEEEhowto:49}
Y. Makris, J. Collins, A. Orailoglu, and P. Vishakantaiah, ''Transparency-based Hierarchical Test Generation for Modular RTL Designs'', Proc. of the 2000 IEEE Intern. Symposium on Circuits and Systems, ISCAS 2000, 28-31 May 2000, Geneva, Switzerland, pp. 689-692.
\bibitem{IEEEhowto:50} R. Kunda, P. Narain, j. Abraham, and B. Rathi, ''Speed up of Test Generation Using High-Level Primitives'', 27h ACM/IEEE Design Automation Conference, DAC, 24-28 June 1990, Orlando, FL, USA, pp. 594-599.
\bibitem{IEEEhowto:51} A. Magdolen, J. Bezakova, E. Gramatova, and M. Fischerova, ''REGGEN – Test Pattern Generation on Register Transfer Level'', Proc. EURO-DAC ‘93, 20-24 Sept. 1993, Hamburg, Germany, pp. 259-264.
\bibitem{IEEEhowto:52}
H. Li, Y. Min, and Z.Li, ''An RT-level ATPG Based on Clustering of Circuit States'', 10th Asian Test Symposium, 19-21 Nov. 2001, Kyoto, Japan, pp. 213-218.
\bibitem{IEEEhowto:53}
T. Hosokawa, R. Inoue, and H. Fujiwara, ''Fault-dependent/independent Test Generation Methods for State Observable FSMs'', 16th Asian Test Symposium, ATS '07, 8-11 Oct. 2007, Beijing, China, pp. 275-280.
\bibitem{IEEEhowto:54}
R. Inoue, T. Hosokawa, and H. Fujiwara, ''A Test Generation Method for State-Observable FSMs to Increase Defect Coverage under the Test Length Constraint'', 17th Asian Test Symposium, ATS '08, 24-27 Nov. 2008, Sapporo, Japan, pp. 27-34.
\bibitem{IEEEhowto:55}
G. Guglielmo, F. Fummi, C. Marconcini, and G. Pravadelli, ''Improving high-level and gate-level testing with FATE: A functional automatic test pattern generator traversing unstabilised extended FSM'', IET Computers \& Digital Techniques, vol. 1, issue 3, May 2007, pp. 187-196.
\bibitem{IEEEhowto:56}
M. Mirzai, M. Tabandeh, B. Alizadeh, and Z. Navabi, ''A New Approach for Automatic Test Pattern Generation in Register Transfer Level Circuits'', IEEE Design \& Test, vol. 30, issue 4, March 2013, pp 49-59.
\bibitem{IEEEhowto:57}
I. Pomeranz, S. Patil, and P. Parvathala, ''A Functional Fault Model with Implicit Fault Effect Propagation Requirements'', 15th Asian Test Symposium 2006, ATS ‘06, 20-.23 Nov. 2006, Fukuoka, Japan, pp. 95-102.
\bibitem{IEEEhowto:58} M. Lajolo, M. Rebaudengo, M. Reorda, M. Violante, ''Behavioral-level Test Vector Generation for System-on-Chip Design'', Proc. IEEE Intern. High-Level Design Validation and Test Workshop, 8-10 Nov. 2000, Berkeley, CA, USA, pp. 21-25.
\bibitem{IEEEhowto:59}
S. Ravi, I.Ghosh, V. Boppana, and K. Jha, ''Fault-Diagnosis-Based Technique for Establishing RTL and Gate-Level Correspondences'', IEEE Trans. on CAD, vol. 20, no. 12, December 2001, pp. 1414-1425.
\bibitem{IEEEhowto:60}
M. Rumplik, and J. Strnadel, ''On RTL Testability and Gate-Level Stuck-At-Fault Coverage Correlation for Scan Circuits'', 14th Euromicro Conf. on Digital System Design, DSD’11, 31 Aug. - 2 Sept. 2011, Oulu, Finland, pp. 367-374.
\bibitem{IEEEhowto:61}
A. Al-Yamani, and E. McCluskey, ''Test Quality for High Level Structural Test'', Ninth IEEE intern. High-level Design Validation and Test Workshop, HLDVT’04, 10-12 Nov. 2004, Sonoma Valley, CA, USA, pp. 109-114.
\bibitem{IEEEhowto:62}
E. Rudnik, F. Corno, R. Vietti, P. Prinetto, A. Ellis, and M. Reorda, ''Fast Sequential Circuit Test Generation Using High-Level and Gate-Level Techniques'', Proc. Design, Automation and Test in Europe, DATE 1998, 23-26 Feb.1998, Paris, France, pp. 570-576.
\bibitem{IEEEhowto:63}
B. Vinnakota, and J. Andrews, ''Fast Fault Translation'', IEEE Trans. on VLSI, vol. 6, no. 1, March 1998, pp. 122-133.
\bibitem{IEEEhowto:64}
M. Prabhu, and J. Abraham, ''Functional Test Generation for Hard to Detect Stuck-At Faults using RTL Model Checking'', 17th IEEE European Test Symposium, ETS, 28-31 May 2012, Annecy France, pp. 1-6.
\bibitem{IEEEhowto:65}
S. Ravi, and N. Jha, ''Fast Test Generation for Circuits with RTL and Gate-Level Views'', Proc. of Intern. Test Conference, 2001, 30 Oct. – 1 Nov., Baltimore, MD, USA, pp. 1068-1077.
\bibitem{IEEEhowto:66} 
C. Chen, C. Wei, T. Lu, and H. Gao, ''Software-Based Self-Testing With Multiple-Level Abstractions for Soft Processor Cores'', IEEE Trans. on VLSI, vol. 15, no. 5, May 2007, pp. 505-517.
\bibitem{IEEEhowto:67}
P. Thaker, V. Agrawal, and M. Zaghloul, ''A Test Evaluation Technique for VLSI Circuits using Register-Transfer Level Fault Modeling'', IEEE Trans. on CAD, vol. 22, issue 8, Aug. 2003, pp. 1104-1113.
\bibitem{IEEEhowto:68}
M. Maniatakos, N. Karimi, C. Tirumurti, A. Jas, and Y. Makris, ''Instruction-Level Impact Comparison of RT- vs. Gate-Level Faults in a Modern Microprocessor Controller'', 27th IEEE VLSI Test Symposium, 3-7 May 2009, Santa Cruz, CA, USA, pp. 9-14.
\bibitem{IEEEhowto:69} J. Campos, and H. Al-Asaad, ''Mutation-Based Validation of High-Level Microprocessor Implementations'', Ninth IEEE intern. High-level Design Validation and Test Workshop, HLDVT’04, 10-12 Nov. 2004, Sonoma Valley, CA, USA, pp. 81-86.
\bibitem{IEEEhowto:70}
C. Hobeika, C. Thibeault, and J. Boland, ''Illegal State Extraction From Register Transfer Level'', 8th IEEE Intern. NEWCAS Conference, 20-23 June 2010, Montreal, QC, Canada, pp. 245-248.
\bibitem{IEEEhowto:71}
C. Hobeika, C. Thibeault, and J. Boland, ''Functional Constraint Extraction From Register Transfer Level for ATPG'', IEEE Trans. on VLSI, vol. 23, no. 2, February 2015, pp. 407-412.
\bibitem{IEEEhowto:72}
C. Chec, and T. Noh, ''VHDL Behavioral ATPG and Fault Simulation of Digital Systems'', IEEE Trans. on Aerospace and Electronic Systems, vol. 34, no. 2, April 1998, pp. 428-447.
\bibitem{IEEEhowto:73}
T. Pecenka, J. Strnadel, Z. Kotasek, and L. Sekanina, ''Testability Estimation Based on Controllability and Observability Parameters'', Proc. of the 9th Euromicro Conf. on Digital System Design, DSD’06, 30 August – 1 September 2006, Dubrovnik, Croatia, pp. 504-514.
\bibitem{IEEEhowto:74}
M. Lynch, and S. Singer, ''A Next Generation Diagnostic ATPG System Using the Verilog HDL'', IEEE Intern. Verilog HDL Conf., 31 March – 2 April 1997, Santa Clara, CA, USA, pp. 56-63.
\bibitem{IEEEhowto:75}
S. Chiusao, F. Corno, P. Prinetto, ''RT-level TPG Exploiting High-Level Synthesis Information'', 17th IEEE VLSI Test Symposium 1999, 25-29 April, 1999, Dana Point, CA, USA, pp. 341-346. 
\bibitem{IEEEhowto:76}
F. Corno, M. Reorda, G. Squillero, ''High-Level Observability for Effective High-Level ATPG'', Proc. of 18th IEEE VLSI Test Symposium 2000, 30 April – 4 May 2000, Montreal, Canada, pp. 411-416.
\bibitem{IEEEhowto:77}
P. Vishakantaiah, J. Abraham, and M. Abadir, ''Automatic Test Knowledge Extraction From VHDL (ATKET)'', 29th ACM/IEEE Design Automation Conf. DAC 1992, 8-12 Jun 1992, Anaheim, CA, USA, pp. 273-278.
\bibitem{IEEEhowto:78}
V. Pitchumani, P. Mayor, and N. Radia, ''Fault Diagnosis using Functional Fault Models for VHDL descriptions'', Proc. Intern. Test Conf. 1991, 26-30 Oct. 1991, Nashville, TN, USA, pp. 327-337.
\bibitem{IEEEhowto:79}
M. Santos, F. Goncalves, I. Teixeira, and J Teixeira, ''Implicit Functionality and Multiple Branch Coverage (IFMB): A Testability Metric for RT-Level'', Proc. Intern. Test Conference, ITC'01, 30 Oct - 1 Nov, Baltimore, MD, USA, pp. 377-385.
\bibitem{IEEEhowto:80} S. Ohtake, M. Inoue, and H. Fujiwara, ''A Method of Test Generation for Weakly Testable Data Paths Using Test Knowledge Extracted from RTL Description'', Proc. of 8th Asian Test Symposium, ATS’99, 18 Nov. 1999, Shanghai, China, pp. 5-12.
\bibitem{IEEEhowto:81}
F. Javaheri, M. Namaki-Shoushtari, P. Kamranfar, and Z. Navabi, ''Mapping Transaction Level Faults to Stuck-at Faults in Communication Hardware'', 20th Asian Test Symposium, ATS '11, 20-23 Nov. 2011, New Delhi, India, pp. 114-119.
\bibitem{IEEEhowto:82} Y. Lin, F. Lu, and K. Cheng, ''Pseudofunctional Testing'', IEEE Trans. on VLSI, vol. 25, no. 8, August 2006, pp. 1535-1546.
\bibitem{IEEEhowto:83} H. Tsai, K. Cheng, and S. Bhawmik, ''Improving The Test Quality for Scan-based BIST using A General Test Application Scheme'', 27th Conf. on Design Automation, DAC, 21-25 June 1999, New Orleans, LA, USA, pp. 748-753.
\bibitem{IEEEhowto:84} H. Tsai, K. Cheng, and S. Bhawmik, ''On Improving Test Quality of Scan-Based BIST'', IEEE Trans. on CAD, vol. 19, no. 8, August 2000, pp. 928-938.
\bibitem{IEEEhowto:85} M. Santos, J. Fernandes, I Teixeira, and J. Teixeira, ''RTL Test Pattern Generation for High Quality Loosely Deterministic BIST'', Proc. of Design, Automation and Test in Europe, DATE 2003, 7th Mar. 2003, Munich, Germany, pp. 994-999.
\bibitem{IEEEhowto:86} D. Berthelot, M. Flottes, and B. Rouzeyre, ''BISTing Data Paths at Behavioral Level'', IEEE Intern. Test Conf., ITC'00, Oct. 2000, Atlantic City, NJ, CA, USA, pp. 672-680.
\bibitem{IEEEhowto:87} T. Masuzawa, M. Izutsu, and H. Wada, ''Single-control Testability of RTL Data Paths for BIST'', Proc. of the Ninth Asian Test Symposium, ATS’00, 6th December 2000, Taipei, Waiwan, pp. 210-215.
\bibitem{IEEEhowto:88} I. Ghosh, N. Jha, and S. Bhawmik, ''A BIST Scheme for RTL Circuits Based on Symbolic Testability Analysis'', IEEE Trans. on CAD, vol. 19, no. 1, August 2000, pp. 111-128.
\bibitem{IEEEhowto:89}
N. Yogi, and V. Agrawal, ''Spectral RTL Test Generation for Gate-Level Stuck-at Faults'', 15th Asian Test Symposium, ATS'06, 20-23 Nov. 2006, Fukuoka, Japan, pp. 83-88.
\bibitem{IEEEhowto:90}
N. Yogi, and V. Agrawal, ''Spectral RTL Test Generation for Microprocessors'', 20th Intern. Conf. on VLSI Design, Jan. 2007, Bangalore, India, pp. 473-478.
\bibitem{IEEEhowto:91} M. Karunaratne, and F. Hill, ''A Vector Based Backward State justification Search for Test Generatio in Sequential Circuits'', Proc. of Ninth Annual Intern. Phoenix Conf. on Computers and Communications, 21-23 march 1990, Scottsdale, AZ, USA, pp. 630-637.
\bibitem{IEEEhowto:92}
S. Ohtake, N. Ikeda, M. Inoue, and H. Fujiwara, ''A Method of Unsensitizable Path Identification using High Level Design Information'', 5th Intern. Conf. on Design and Technology of Integrated Systems in Nanoscale Era, DTIS 2010, 23-25 March 2010, Hammamet, Tunesia, pp. 1-6.
\bibitem{IEEEhowto:93}
T. Strauch, "Deriving AOC C-Models from D\&V Languages for Single- or Multi-Threaded Execution Using C or C++", 18. Workshop Methoden und Beschreibungssprachen zur Modellierung und Verifikation von Schaltungen und Systemen, MBMV 2015, 3-4 March, Chemnitz, Germany, pp. 173-182. 
\bibitem{IEEEhowto:94}
F. Ferrandi, G. Ferrara, D. Sciuto, A. Fin, and F. Fummi, ''Functional Test Generation for Behaviorally Sequential Models'', Proc. of Design, Automation and Test in Europe, DATE 2001, 13-16 Mar. 2001, Munich, Germany, pp. 403-410.


\end{thebibliography}
%

%

\begin{IEEEbiographynophoto}{Tobias Strauch}
 received his Diploma (FH) at the University of applied science (FH) Furtwangen, Germany in '98. He works for EDAptix in Munich, Germany. His field of interests are hardware assisted verification, PDVL, TLM, C-Slow Retiming, System Hyper Pipelining, High Level ATPG, FPGA debugging and wave based data transfer.
\end{IEEEbiographynophoto}




\end{document}